\documentclass[aps,twocolumn,prl,showpacs,showkeys,preprintnumbers,superscriptaddress,nobibnotes,floatfix,longbibliography,notitlepage,nofootinbib]{revtex4-2}

\pdfoutput=1
\usepackage{amsmath,amssymb}
\usepackage{amsfonts}
\usepackage{bm,bbm}
\usepackage{graphicx}
\usepackage{mathrsfs}
\usepackage{slashed}
\usepackage{booktabs}
\usepackage{tabu}
\usepackage[dvipsnames]{xcolor}
\usepackage[normalem]{ulem}
\usepackage{soul}
\usepackage{todonotes}

\usepackage{hyperref}
\hypersetup{colorlinks,citecolor= blue,linkcolor= blue, urlcolor=blue}

\usepackage{xcolor}
\usepackage{ulem}
\usepackage{array}
\usepackage{verbatim}
\usepackage{epsfig}
\usepackage{multirow}
\usepackage{amsmath}

\newcommand{\be}{\begin{equation}}
\newcommand{\ee}{\end{equation}}
\newcommand{\ba}{\begin{array}}
\newcommand{\ea}{\end{array}}
\newcommand{\bea}{\begin{eqnarray}}
\newcommand{\eea}{\end{eqnarray}}

\usepackage{ulem,fancyvrb}
\usepackage{xcolor} 
\usepackage{cleveref}

\begin{document}

\title{A New Probe for Long-Lived Particles at Higgs Factories: Displaced Photons in the Hadronic Calorimeter}

\author{Zhicheng Jiang}
\email{zhicheng.jiang@m.scnu.edu.cn}
\author{Hengne Li\mbox{\,\href{https://orcid.org/0000-0002-2366-9554}{\includegraphics[scale=0.075]{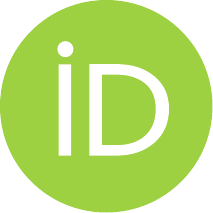}}} }
\email{hengne.li@m.scnu.edu.cn}
\author{Jin-Han Liang\mbox{\,\href{https://orcid.org/0000-0002-6141-216X}{\includegraphics[scale=0.075]{orcid.pdf}}} }
\email{jinhanliang@m.scnu.edu.cn}

\affiliation{State Key Laboratory of Nuclear Physics and
Technology, Institute of Quantum Matter, South China Normal
University, Guangzhou 510006, China}
\affiliation{Guangdong Basic Research Center of Excellence for
Structure and Fundamental Interactions of Matter, Guangdong
Provincial Key Laboratory of Nuclear Science, Guangzhou
510006, China}

\begin{abstract}
The search for dark matter and other photon-portal long-lived particles (LLPs) at electron-positron colliders often relies on the mono-photon signature. 
At future Higgs factories operating at the $Z$-pole, this approach faces a critical challenge: the irreducible background from $e^+e^- \to \nu\bar{\nu}\gamma$ becomes overwhelming. 
We propose a novel strategy that overcomes this limitation by searching for displaced photons from LLP decays within the barrel of the hadronic calorimeter. 
This signature exploits the architectural  shielding of the detector to create a nearly background-free environment. 
Our analysis demonstrates exceptional sensitivity to LLPs with decay lengths from $\sim$1 to $10^6$ meters, improving upon conventional searches by up to one order of magnitude for benchmark photon-portal models.
\end{abstract}

\maketitle 

\section*{Introduction}
The particle nature of dark matter (DM) and the possible existence of a dark sector (DS) stand among the central challenges in fundamental physics~\cite{Essig:2013lka,Bozorgnia:2024pwk}.
Decades of searches for weakly interacting massive particles (WIMP) have yielded no positive signals and imposed increasingly stringent limits~\cite{Roszkowski:2017nbc,Schumann:2019eaa}. The broader landscape of feebly interacting particles, such as light DS states interacting with ordinary particles via portals, presents unique experimental opportunities \cite{Gori:2022vri}.
Electron-positron colliders offer an ideal environment for such searches, providing known initial states and clean event topologies.

A crucial signature in this search is the mono-photon event, characterized by a single, energetic photon with no other detectable particles in the detector.
This signature serves as a direct portal to new physics involving invisible final states~\cite{Birkedal:2004xn, Fox:2011fx, Essig:2013vha}.
Its sensitivity extends beyond stable DM candidates~\cite{Fox:2011fx, Essig:2013vha, Dolan:2017osp, Belle-II:2018jsg, Liu:2019ogn, Liang:2021kgw, Liu:2018jdi, Liang:2019zkb}
to a broad range of models, in which unstable DS states interact with the Standard Model (SM) through photon portals.
Prominent examples include the dark axion portal~\cite{deNiverville:2018hrc, Zhevlakov:2022vio, Jodlowski:2024lab}, models featuring a neutrino dipole moment~\cite{Magill:2018jla, Zhang:2023nxy, Jodlowski:2024lab}, inelastic DM scenarios with dipole operators~\cite{Izaguirre:2015zva, Dienes:2023uve, Barducci:2024nvd}, and neutralinos \cite{OPAL:1990xgw}.

\begin{figure}[htbp]
\centering
\includegraphics[width=\linewidth]{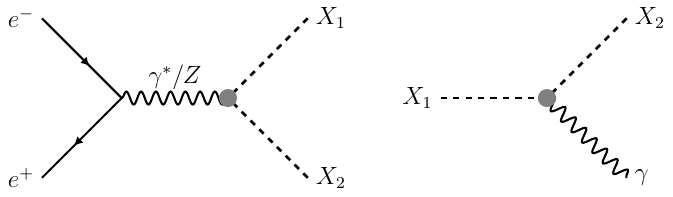}
\caption{Feynman diagrams of the production (left) and decay (right) processes of photon-portal DS particles at Higgs factories, leading to the mono-photon signature.
}
\label{fig:feyn-diag}
\end{figure}

These photon-portal models share a common phenomenological feature, where a DS particle $X_1$ undergoes radiative decay $X_1 \to \gamma X_2$, with $X_2$ representing an invisible particle such as a SM neutrino or a lighter DS state. 
The canonical search strategy at electron-positron colliders targets the process $e^+e^- \to X_1 X_2$ followed by this decay, which yields a photon pointing back to the interaction point, as depicted in Fig.~\ref{fig:feyn-diag}~\cite{Zhang:2023nxy, Jodlowski:2024lab}.

However, the unprecedented integrated luminosities planned at future Higgs factories, specifically the CEPC~\cite{CEPCStudyGroup:2025kmw,CEPCStudyGroup:2023quu,CEPCStudyGroup:2018rmc,CEPCStudyGroup:2018ghi} and FCC-ee~\cite{FCC:2018evy,Agapov:2022bhm} operating at the $Z$-pole, introduce a fundamental experimental challenge. 
The irreducible SM background from $e^+e^- \to \nu\bar{\nu}\gamma$ becomes overwhelming, with $\mathcal{O}(10^{10})$ expected events for photons above 1\,GeV at 100\,ab$^{-1}$ integrated luminosity. 
This background severely limits sensitivity to new photon-portal scenarios.

When photon-portal couplings are sufficiently suppressed, $X_1$ emerges as a long-lived particle (LLP), offering potential avenues to circumvent prompt backgrounds through displaced decays. 
One promising technique is the non-pointing photon signature~\cite{ATLAS:2013etx,Duarte:2023tdw,Beltran:2024twr}. The main challenge in identifying these photons within the electromagnetic calorimeter (ECAL) is the immense rate of indistinguishable background events, which requires exceptionally high granularity to determine their direction of origin.

In this Letter, we propose a transformative approach that fundamentally redefines this background limitation. 
We demonstrate that the hadronic calorimeter (HCAL), traditionally reserved for jet energy measurement, can be repurposed as an ideal far detector for photons from the decay of LLPs. This is achievable provided that a high granular HCAL design is implemented.

\begin{figure}[htbp]
\centering
\includegraphics[width=\linewidth]{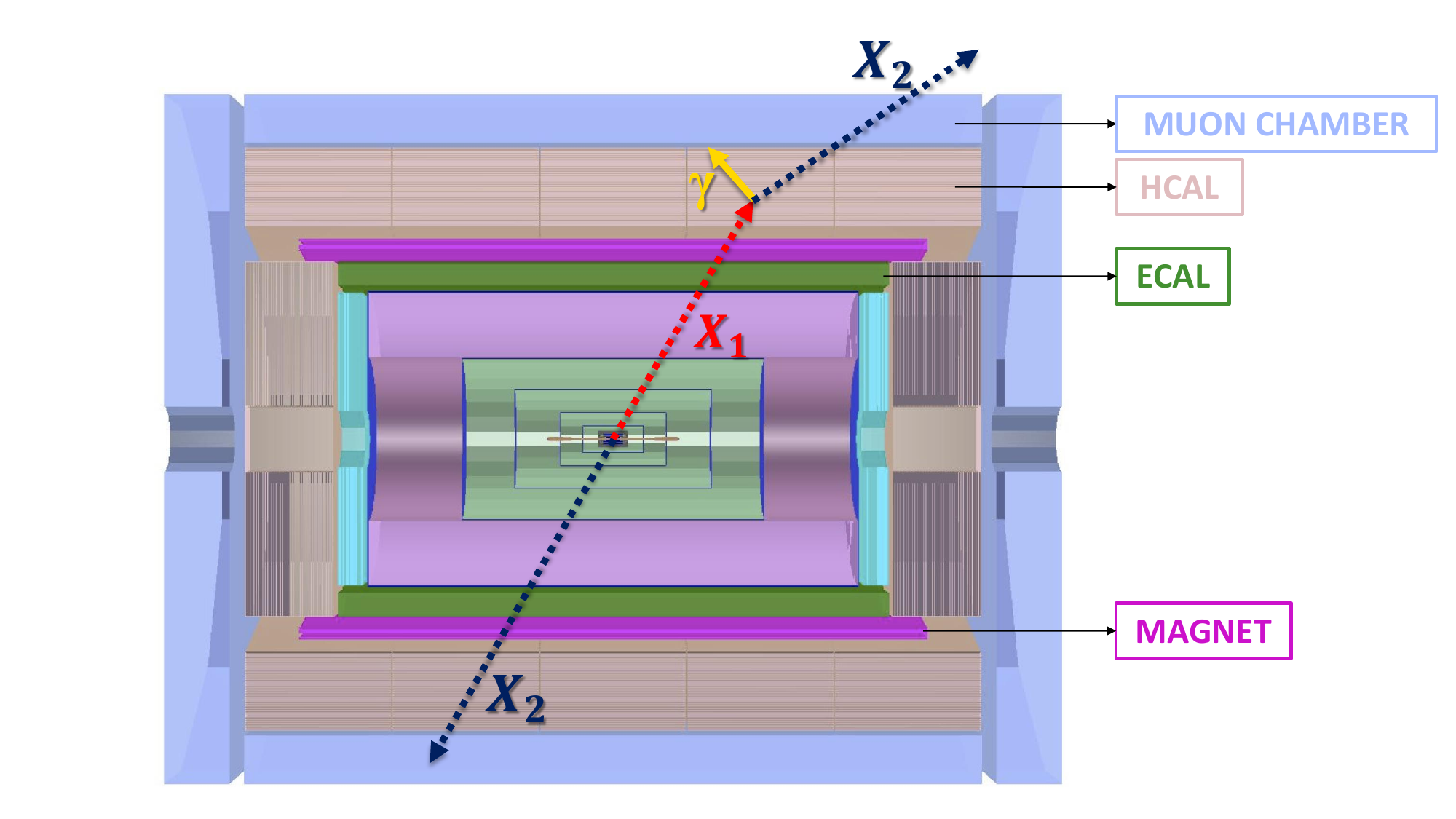}
\put(-180,-10){(a) HCAL-$\gamma_d$ signature}
\vspace{2mm}
\includegraphics[width=\linewidth]{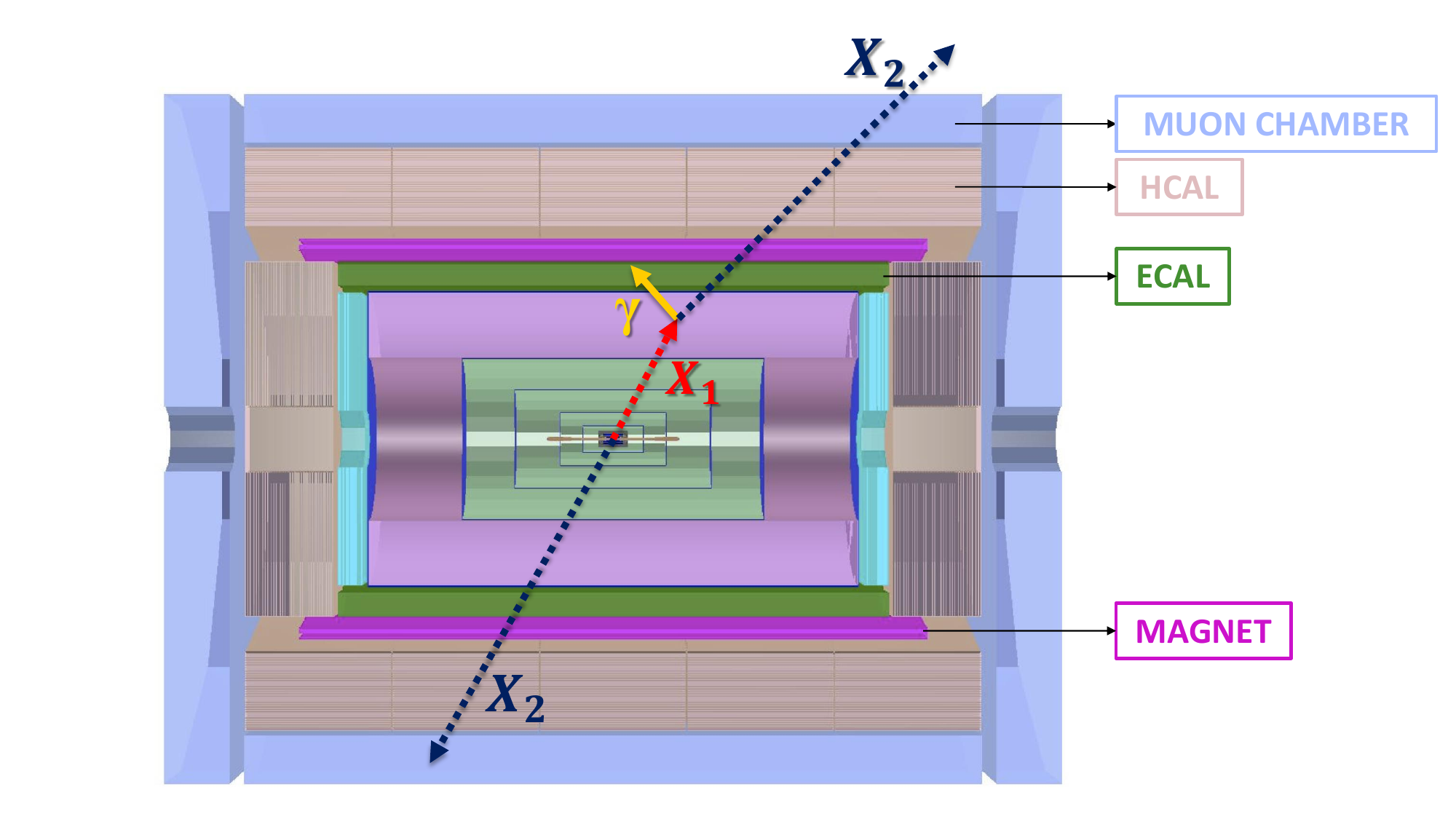}
\put(-180,-10){(b) ECAL-$\gamma_d$ signature}
\vspace{2mm}
\includegraphics[width=\linewidth]{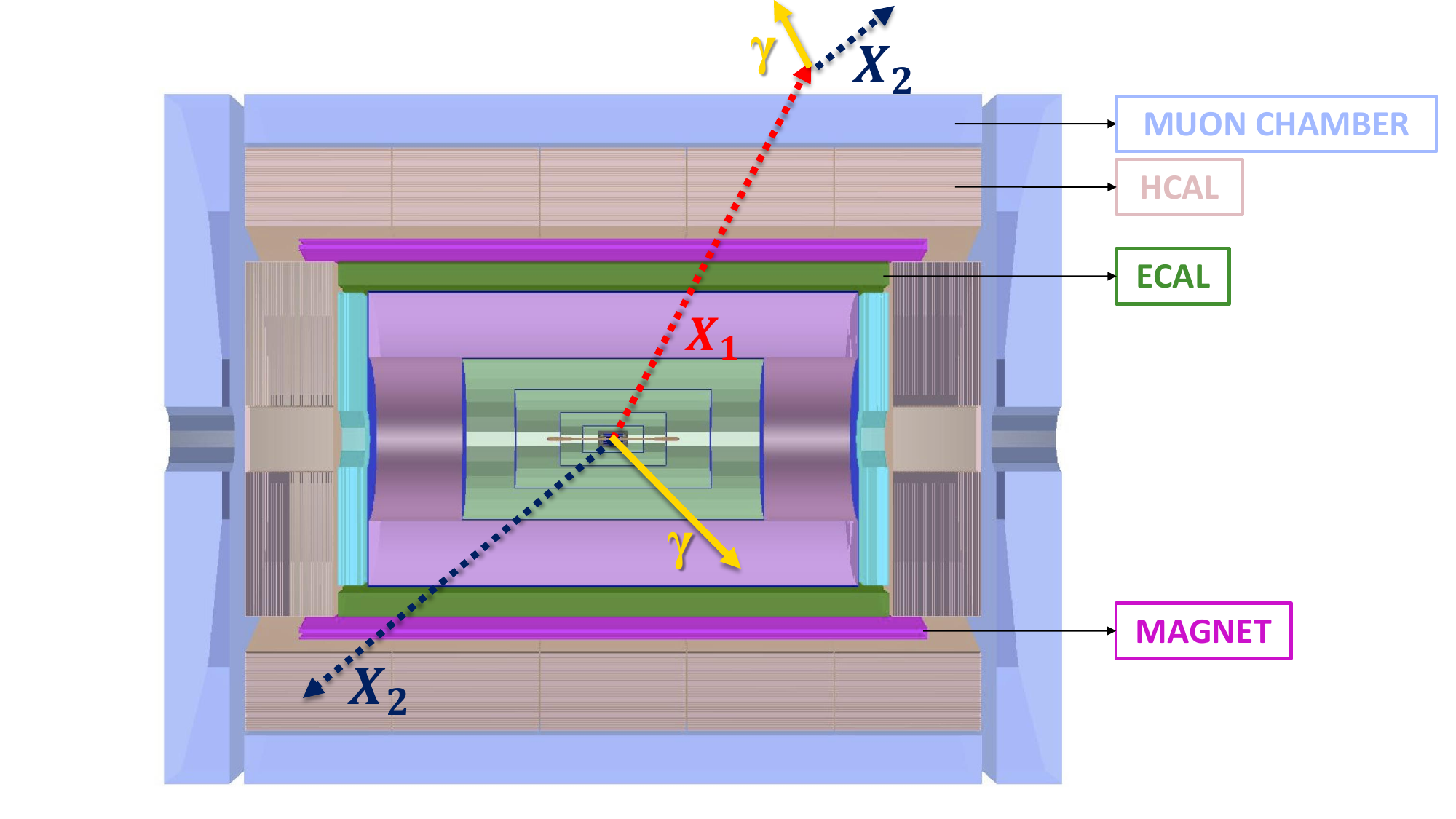}
\put(-180,-10){(c) ECAL-$\gamma_r$ signature}
\caption{Schematic diagrams for the three mono-photon signatures at CEPC Reference Detector, including (a) HCAL-$\gamma_d$ signature, (b) ECAL-$\gamma_d$ signature, and (c) ECAL-$\gamma_r$ signature, where $\gamma_r$ and $\gamma_d$ denote photons generated by initial state radiation and the radiative decay of DS particles, respectively.
Here, $X_2$ and $X_1$ are DS particles generated by $e^+ e^- \to X_2 X_1 (\gamma_r)$, with subsequent decay of $X_1 \to X_2 \gamma_d$.
}
\label{fig:channels}
\end{figure}

We introduce the HCAL-$\gamma_d$ signature, depicted in Fig.~\ref{fig:channels} (a), where $X_1$ is produced via $e^+e^- \to X_1 X_2$, traverses the inner tracker and ECAL undecayed, and decays within the HCAL barrel. 
The resulting displaced electromagnetic shower can be efficiently reconstructed using modern particle flow algorithms (PFA)~\cite{PandoraPFA2009, Ruan:2013rkk, Ruan:2018yrh} for a high-granularity HCAL. 
Crucially, this signature inherently exploits the hermeticity of the detector, as the surrounding ECAL, HCAL endcaps, and muon system collectively provide powerful veto capabilities that filter out standard prompt photons and charged particles.

Our comprehensive analysis reveals that this HCAL-$\gamma_d$ signature provides exceptional sensitivity to photon-portal LLPs across an extensive range of decay lengths, spanning from $\sim 1$ to $10^6$ meters. 
For benchmark models, this approach improves sensitivity by factors of 10–20 compared to conventional searches, opening a new and powerful window into long-lived DS phenomena.

\section*{The HCAL-$\gamma_d$ signature}
Our analysis utilizes the proposed CEPC Reference Detector (Ref-Det)~\cite{CEPCStudyGroup:2025kmw} as a benchmark experimental setup. The detector features a conventional $4\pi$ geometry comprising a central tracking system, ECAL, HCAL, and muon detection system, achieving nearly full angular coverage of $|\cos\theta| < 0.99$.

The HCAL barrel serves as our fiducial far detector for displaced photon signatures from LLP decays. It spans 1315 mm radially with inner radius $R_{\text{in}}^H = 2140$ mm and outer radius $R_{\text{out}}^H = 3455$ mm, covering the pseudorapidity range $|\eta| < 1.2$ ($33.5^\circ < \theta < 146.5^\circ$). The baseline glass scintillator (GS) HCAL design features a 48-layer structure providing approximately $6\lambda_I$ of hadronic absorption.
The radiation length per layer in GS-HCAL is $1.44X_0$, compared to $1.21X_0$ in the plastic scintillator (PS) alternative.

The fine granularity of the GS-HCAL, with cell size $40 \times 40 \times 10$ mm$^3$, enables effective discrimination between photon-induced electromagnetic showers and neutral hadronic interactions through the PFA. This capability is critical for background suppression in LLP searches.

In our analysis, displaced photons in the HCAL barrel are assumed to be efficiently reconstructed via PFA, with a baseline criterion of energy deposition in at least five active layers and a reconstruction efficiency of 50\%.
This five-layer requirement leads to a photon energy threshold $E_\gamma^{\text{th}}$ of 2.3 GeV (1.37) GeV for GS-HCAL (PS-HCAL) according to 
$E_\gamma^{\text{th}} = E_c 2^t$~\cite{Grupen:2008zz},
where $E_c$ is the critical energy of the HCAL with 15.7 (21.7) MeV for GS-HCAL (PS-HCAL) and $t$ is the total radiation length for the five-layer structure.
A more precise determination of the energy threshold and the reconstruction efficiency using detailed \textsc{GEANT4} simulation is deferred to future work.

Although the PS-HCAL design achieves a lower threshold due to its smaller radiation length per layer, the GS-HCAL configuration provides a substantially higher sampling fraction of approximately 31\%, far exceeding the 1.6\% offered by the PS-HCAL option. This enhanced sampling ratio significantly improves energy resolution and detection efficiency for electromagnetic showers, providing important compensating advantages despite the higher energy threshold.
Hence, we adopt the GS-HCAL as the benchmark HCAL in this work.

Crucially, the HCAL barrel is fully enclosed by the ECAL barrel (employing 300 mm of BGO crystals corresponding to $24X_0$ and $1.35\lambda_I$), HCAL endcaps (providing $50X_0$, $6\lambda_I$ of additional shielding), and the muon chamber system (6 layers totaling $40X_0$, $4.2\lambda_I$ of absorber material). This configuration ensures a clean, well-isolated environment for identifying decays within the HCAL barrel volume, with excellent suppression against external backgrounds.

The dominant backgrounds for the HCAL-$\gamma_d$ signature originate from neutral particles produced at the primary vertex that evade detection in the inner tracker and ECAL, subsequently depositing energy in the HCAL barrel. These can be categorized into three classes: photon-induced, neutrino-induced, and hadron-induced backgrounds.
Combining all sources, the total expected background for the HCAL-$\gamma_d$ signature is $N_b \sim 10^2$–$10^3$ events across the full 100 ab$^{-1}$ dataset. 
The details of the background estimation are provided in the Supplemental Materials.
The dominant uncertainty stems from hadron-induced backgrounds due to limited simulation statistics and the imperfectly known efficiency of distinguishing photons from neutral hadrons in the HCAL.
To account for these uncertainties, we perform sensitivity calculations with both conservative ($N_b = 10^4$) and optimistic ($N_b = 10^2$) background estimates. This background level represents a reduction of over six orders of magnitude compared to the $\mathcal{O}(10^{10})$ $e^+e^- \to \nu\bar{\nu}\gamma$ events that plague conventional ECAL-based mono-photon searches, underscoring the transformative potential of the HCAL-$\gamma_d$ approach.

For comparison, we also evaluate the sensitivities of two conventional ECAL-based mono-photon searches~\cite{deNiverville:2018hrc, Zhevlakov:2022vio, Jodlowski:2024lab, Magill:2018jla, Zhang:2023nxy}, referred to as ECAL-$\gamma_d$ from DS particle decays, and ECAL-$\gamma_r$ from initial state radiation (ISR) process.
Together with our proposed HCAL-$\gamma_d$ signature, these channels provide complementary coverage across different LLP decay regimes. 
Their defining features are summarized in Table~\ref{tab:monophoton_simple}, with corresponding schematic diagrams shown in Fig.~\ref{fig:channels}.
Further details of the signal region and SM background analysis for the ECAL-based monophoton signatures are given in the Supplemental Materials.
The complementary nature of these signatures ensures comprehensive coverage across the parameter space of photon-portal models, with each channel dominating in specific decay length regimes.

\begin{table}[h!]
\centering
\caption{Three mono-photon signatures at Higgs factories.}
\label{tab:monophoton_simple}
\begin{tabular}{lll}
\textbf{Signature} & \textbf{Photon origin} & \textbf{Decay in detector} \\
\hline
HCAL-$\gamma_d$ & $X_1$ decay & HCAL Barrel \\
ECAL-$\gamma_d$ & $X_1$ decay & ECAL \\
ECAL-$\gamma_r$ & ISR & ECAL \\
\hline
\end{tabular}
\end{table}

\section*{Benchmark Model}
We employ the dark axion portal as our primary benchmark scenario to demonstrate the efficacy of the HCAL-$\gamma_d$ signature. This framework provides an interaction that enables simultaneous coupling between Standard Model particles and both dark photons and dark axions. The relevant interaction is described by the dimension-5 operator~\cite{Kaneta:2016wvf}:

\begin{equation}
\mathcal{O}_{a\gamma'B} \equiv g_B a \tilde{F}_{\mu\nu}^\prime B^{\mu\nu},
\end{equation}
where $a$ denotes the axion-like particle (ALP), $F_{\mu\nu}^\prime$ represents the field-strength tensor of the dark photon $\gamma^\prime$, $B^{\mu\nu}$ is the hypercharge field-strength tensor, and $g_B$ is the coupling coefficient with mass dimension $-1$.

Assuming the dark photon is heavier than the dark axion, we identify $\gamma^\prime$ with $X_1$ and the ALP with $X_2$ as shown in Fig.~\ref{fig:channels}. The decay width for $\gamma^\prime \to a\gamma$ is given by~\cite{Kaneta:2016wvf}:
\begin{equation}
\Gamma_{\gamma^\prime \to a\gamma} = \frac{1}{24\pi}g_B^2 c_W^2 m_{\gamma^\prime}^3 (1-r_m^2)^3,
\label{eq:decay_width}
\end{equation}
where $r_m = m_a/m_{\gamma^\prime}$, with $m_a$ and $m_{\gamma^\prime}$ denoting the masses of the ALP and dark photon respectively, and $c_W = \cos\theta_W$ is the cosine of the Weinberg angle.

The interaction with the hypercharge field induces couplings to both the SM photon and $Z$ boson. For CEPC operating at the $Z$-pole, the dominant production process is $e^+e^- \to Z \to a\gamma^\prime$. We consider the regime $m_{\gamma^\prime} - m_a < m_Z$ to ensure the radiative decay dominates.

The photon energy spectrum from dark photon decay follows a flat distribution in the laboratory frame, with energies bounded by:
\begin{equation}
E_\gamma^{\text{max, min}} = E_\gamma^* \gamma (1 \pm \beta),
\label{eq:photon_energy}
\end{equation}
where $E_\gamma^* = \frac{1}{2} m_{\gamma'}(1-r_m^2)$ is the photon energy in the dark photon rest frame, $\gamma = E_{\gamma'}/m_{\gamma'}$ is the Lorentz boost factor, and $\beta = \sqrt{1-1/\gamma^2}$ is the corresponding velocity.

\section*{Results and Findings}
\begin{figure*}[htbp]
\centering
\includegraphics[width=0.42\textwidth]{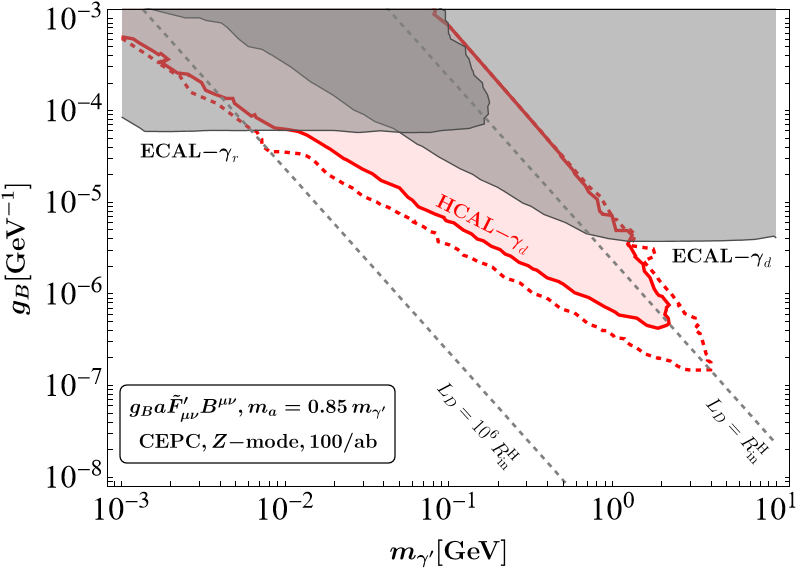}
\put(-180,66){\bf (a)}
\hspace{5mm}
\includegraphics[width=0.42\textwidth]{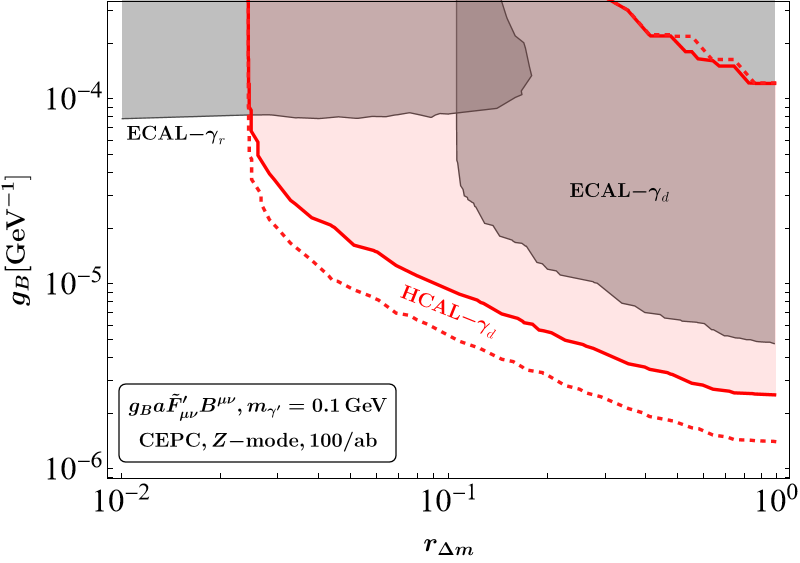}
\put(-180,66){\bf (b)}
\caption{
Projected $2\sigma$ sensitivity contours for the axion portal model at the CEPC $Z$-mode run. 
\textbf{(a)} Contours in the plane of coupling ($g_B$) versus dark photon mass ($m_{\gamma^\prime}$), with the axion mass fixed at $m_a = 0.85 m_{\gamma^\prime}$. 
The sensitivities are shown for several mono-photon signatures, including ECAL-$\gamma_d$, ECAL-$\gamma_r$, and HCAL-$\gamma_d$.
For the HCAL-$\gamma_d$ signature, the solid (dashed) curves correspond to conservative (optimistic) background estimates with $N_b = 10^4$ ($10^2$).
Gray dashed lines indicate reference decay length $L_D = R_{\rm in}^H$ and $L_D = 10^6\, R_{\rm in}^H$ for the dark photon in the laboratory frame.
\textbf{(b)} Contours as a function of mass-splitting parameter $r_{\Delta m } \equiv (m_{\gamma'}-m_a)/m_{\gamma'}$ for fixed dark photon mass $m_{\gamma^\prime} = 0.1\,\rm GeV$. 
}
\label{fig:con-AP}
\end{figure*}

We begin by quantifying the discovery potential of the HCAL-$\gamma_d$ signature using the dark axion portal as a benchmark scenario. Fig.~\ref{fig:con-AP}(a) presents the projected $2\sigma$ sensitivity contours in the plane of coupling strength $g_B$ versus dark photon mass $m_{\gamma'}$, with the axion mass fixed at $m_a = 0.85 m_{\gamma'}$. Our analysis focuses on the CEPC $Z$-pole operation at $\sqrt{s} = 91.2$ GeV with an integrated luminosity of $\mathcal{L} = 100$ ab$^{-1}$.

The HCAL-$\gamma_d$ channel demonstrates remarkable sensitivity across the intermediate decay length regime ($R_{\text{in}}^H \lesssim L_D \lesssim 10^6 R_{\text{in}}^H$), achieving $g_B \sim 5\times10^{-7}$ GeV$^{-1}$ for conservative background estimates ($N_b = 10^4$) and $g_B \sim 2 \times 10^{-7}$ GeV$^{-1}$ for optimistic estimates ($N_b = 10^2$) for GeV-scale dark photons. This represents a substantial improvement of 10–20$\times$ over the conventional ECAL-$\gamma_d$ channel.

The three mono-photon signatures exhibit beautifully complementary behavior: the ECAL-$\gamma_d$ channel dominates for short decay lengths ($L_D \lesssim R_{\text{in}}^H$), the HCAL-$\gamma_d$ signature excels in the intermediate regime, and the ECAL-$\gamma_r$ channel provides coverage for ultra-long-lived particles ($L_D \gtrsim 10^6 R_{\text{in}}^H$). The HCAL-$\gamma_d$ signature uniquely bridges the critical gap between prompt and ultra-long-lived regimes that has remained largely inaccessible to previous search strategies.

Fig.~\ref{fig:con-AP}(b) explores the sensitivity dependence on mass splitting for fixed dark photon mass $m_{\gamma'} = 0.1$ GeV. We define the mass-splitting parameter $r_{\Delta m} \equiv (m_{\gamma'}-m_a)/m_{\gamma'}$ to characterize this behavior.
For $ 0.04 \lesssim  r_{\Delta m } \lesssim 0.1$, the HCAL-$\gamma_d$ signature outperforms the ECAL mono-photon signatures by roughly an order of magnitude.
The HCAL-$\gamma_d$ sensitivity exhibits a pronounced dependence on $r_{\Delta m}$, diminishing significantly when $r_{\Delta m} \lesssim 0.02$. This threshold corresponds to the point where the maximum photon energy $E_\gamma^{\text{max}}$ from dark photon decay falls below our detection threshold $E_\gamma^{\text{th}}$.
This limitation could potentially be mitigated through improved reconstruction algorithms that require fewer active layers, though such optimization is deferred to future studies.
The ECAL-$\gamma_d$ signature also loses sensitivity when $r_{\Delta m} \lesssim 0.1$ due to the energy cut imposed to suppress QED backgrounds; see the Supplemental Materials for details.

The exceptional performance of the HCAL-$\gamma_d$ channel arises from its intrinsically clean environment, enabling unprecedented signal-to-background sensitivity at intermediate decay lengths where the geometric acceptance remains large and the background negligible.
The combination of large decay volume, efficient photon reconstruction, and powerful background rejection creates an ideal discovery environment for photon-portal LLPs.
To highlight the broader applicability of the proposed HCAL-$\gamma_d$ signature, we also present sensitivities to the neutrino dipole portal operator in the Supplemental Materials.

\section*{Summary and Outlook}
In this Letter, we have established the HCAL-$\gamma_d$ signature as a radical approach for discovering LLPs at future electron-positron colliders. By reconceptualizing the HCAL as a far detector for displaced photon from DS particle decays, we have turned the fundamental challenge of immense prompt backgrounds into a powerful discovery opportunity.

Our comprehensive analysis demonstrates that the HCAL-$\gamma_d$ signature provides exceptional sensitivity to photon-portal LLPs across an unprecedented range of decay lengths, spanning six orders of magnitude. This extensive coverage bridges a crucial gap between conventional prompt and ultra-long-lived search strategies that has remained largely inaccessible to previous approaches.

The core strength of this method lies in its exploitation of the inherent hermeticity of the detector. The shielding provided by the surrounding ECAL, HCAL endcaps, and muon system creates a veto-rich environment that filters out standard prompt backgrounds, resulting in a nearly background-free search channel. Our detailed background analysis confirms the expected total background reduction of over six orders of magnitude compared to conventional mono-photon searches.

For benchmark photon-portal models, the HCAL-$\gamma_d$ approach improves sensitivity by over one order of magnitude compared to established search strategies for GeV-scale DS particles. This enhancement is particularly pronounced in the intermediate decay length regime ($R_{\text{in}}^H \lesssim L_D \lesssim 10^6 R_{\text{in}}^H$), where the method achieves its optimal combination of large geometric acceptance and powerful background rejection.

The complementary nature of the three mono-photon signatures provides comprehensive coverage across the full parameter space of photon-portal models. Future experimental programs at Higgs factories should incorporate all three channels to maximize discovery potential.

Looking forward, several directions promise further enhancements. More sophisticated particle identification algorithms could improve the discrimination between photon and hadron showers in the HCAL. Optimized detector designs with enhanced HCAL granularity could lower energy thresholds and improve reconstruction efficiency. Additionally, the HCAL-$\gamma_d$ concept could be extended to other experimental contexts, including hadron colliders and fixed-target experiments.

This work not only introduces a powerful new search strategy but also demonstrates the value of creative detector usage in expanding the reach of particle physics. The HCAL-$\gamma_d$ signature promises to be an essential component of the physics program at future electron-positron colliders, opening a new window into the hidden sector and bringing us closer to understanding the fundamental nature of dark matter.

\section*{Acknowledgments}

We thank our colleagues from the CEPC collaboration for fruitful discussions. This work was supported by the National Natural Science Foundation of China (Grant No. NSFC-12347121) and the National Key R\&D Program (Grant No. 2024YFA1610600) from the Ministry of Science and Technology of China.

\bibliography{ref.bib}{}

\providecommand{\href}[2]{#2}\begingroup\raggedright\begin{thebibliography}{10}

\bibitem{Essig:2013lka}
R.~Essig {\em et~al.}, ``{\it {Working Group Report: New Light Weakly Coupled
  Particles}},'' in {\em {Snowmass 2013}: {Snowmass on the Mississippi}}.
\newblock 10, 2013.
\newblock [\href{http://arxiv.org/abs/1311.0029}{{\ttfamily arXiv:1311.0029}}
  [hep-ph]].

\bibitem{Bozorgnia:2024pwk}
N.~Bozorgnia, J.~Bramante, J.~M. Cline, D.~Curtin, D.~McKeen, D.~E. Morrissey,
  A.~Ritz, S.~Viel, A.~C. Vincent, and Y.~Zhang, ``{\it {Dark matter candidates
  and searches}},'' \href{http://dx.doi.org/10.1139/cjp-2024-0128}{{\em Can. J.
  Phys.} {\bfseries 103} no.~8, (2025) 671--703},
  [\href{http://arxiv.org/abs/2410.23454}{{\ttfamily arXiv:2410.23454}}
  [hep-ph]].

\bibitem{Roszkowski:2017nbc}
L.~Roszkowski, E.~M. Sessolo, and S.~Trojanowski, ``{\it {WIMP dark matter
  candidates and searches{\textemdash}current status and future prospects}},''
  \href{http://dx.doi.org/10.1088/1361-6633/aab913}{{\em Rept. Prog. Phys.}
  {\bfseries 81} no.~6, (2018) 066201},
  [\href{http://arxiv.org/abs/1707.06277}{{\ttfamily arXiv:1707.06277}}
  [hep-ph]].

\bibitem{Schumann:2019eaa}
M.~Schumann, ``{\it {Direct Detection of WIMP Dark Matter: Concepts and
  Status}},'' \href{http://dx.doi.org/10.1088/1361-6471/ab2ea5}{{\em J. Phys.
  G} {\bfseries 46} no.~10, (2019) 103003},
  [\href{http://arxiv.org/abs/1903.03026}{{\ttfamily arXiv:1903.03026}}
  [astro-ph.CO]].

\bibitem{Gori:2022vri}
S.~Gori {\em et~al.}, ``{\it {Dark Sector Physics at High-Intensity
  Experiments}},'' [\href{http://arxiv.org/abs/2209.04671}{{\ttfamily
  arXiv:2209.04671}} [hep-ph]].

\bibitem{Birkedal:2004xn}
A.~Birkedal, K.~Matchev, and M.~Perelstein, ``{\it {Dark matter at colliders: A
  Model independent approach}},''
  \href{http://dx.doi.org/10.1103/PhysRevD.70.077701}{{\em Phys. Rev. D}
  {\bfseries 70} (2004) 077701},
  [\href{http://arxiv.org/abs/hep-ph/0403004}{{\ttfamily
  arXiv:hep-ph/0403004}}].

\bibitem{Fox:2011fx}
P.~J. Fox, R.~Harnik, J.~Kopp, and Y.~Tsai, ``{\it {LEP Shines Light on Dark
  Matter}},'' \href{http://dx.doi.org/10.1103/PhysRevD.84.014028}{{\em Phys.
  Rev. D} {\bfseries 84} (2011) 014028},
  [\href{http://arxiv.org/abs/1103.0240}{{\ttfamily arXiv:1103.0240}}
  [hep-ph]].

\bibitem{Essig:2013vha}
R.~Essig, J.~Mardon, M.~Papucci, T.~Volansky, and Y.-M. Zhong, ``{\it
  {Constraining Light Dark Matter with Low-Energy $e^+e^-$ Colliders}},''
  \href{http://dx.doi.org/10.1007/JHEP11(2013)167}{{\em JHEP} {\bfseries 11}
  (2013) 167}, [\href{http://arxiv.org/abs/1309.5084}{{\ttfamily
  arXiv:1309.5084}} [hep-ph]].

\bibitem{Dolan:2017osp}
M.~J. Dolan, T.~Ferber, C.~Hearty, F.~Kahlhoefer, and K.~Schmidt-Hoberg, ``{\it
  {Revised constraints and Belle II sensitivity for visible and invisible
  axion-like particles}},''
  \href{http://dx.doi.org/10.1007/JHEP12(2017)094}{{\em JHEP} {\bfseries 12}
  (2017) 094}, [\href{http://arxiv.org/abs/1709.00009}{{\ttfamily
  arXiv:1709.00009}} [hep-ph]]. [Erratum: JHEP 03, 190 (2021)].

\bibitem{Belle-II:2018jsg}
{\bfseries Belle-II} Collaboration, W.~Altmannshofer {\em et~al.}, ``{\it {The
  Belle II Physics Book}},'' \href{http://dx.doi.org/10.1093/ptep/ptz106}{{\em
  PTEP} {\bfseries 2019} no.~12, (2019) 123C01},
  [\href{http://arxiv.org/abs/1808.10567}{{\ttfamily arXiv:1808.10567}}
  [hep-ex]]. [Erratum: PTEP 2020, 029201 (2020)].

\bibitem{Liu:2019ogn}
Z.~Liu, Y.-H. Xu, and Y.~Zhang, ``{\it {Probing dark matter particles at
  CEPC}},'' \href{http://dx.doi.org/10.1007/JHEP06(2019)009}{{\em JHEP}
  {\bfseries 06} (2019) 009},
  [\href{http://arxiv.org/abs/1903.12114}{{\ttfamily arXiv:1903.12114}}
  [hep-ph]].

\bibitem{Liang:2021kgw}
J.~Liang, Z.~Liu, and L.~Yang, ``{\it {Probing sub-GeV leptophilic dark matter
  at Belle II and NA64}},''
  \href{http://dx.doi.org/10.1007/JHEP05(2022)184}{{\em JHEP} {\bfseries 05}
  (2022) 184}, [\href{http://arxiv.org/abs/2111.15533}{{\ttfamily
  arXiv:2111.15533}} [hep-ph]].

\bibitem{Liu:2018jdi}
Z.~Liu and Y.~Zhang, ``{\it {Probing millicharge at BESIII via monophoton
  searches}},'' \href{http://dx.doi.org/10.1103/PhysRevD.99.015004}{{\em Phys.
  Rev. D} {\bfseries 99} no.~1, (2019) 015004},
  [\href{http://arxiv.org/abs/1808.00983}{{\ttfamily arXiv:1808.00983}}
  [hep-ph]].

\bibitem{Liang:2019zkb}
J.~Liang, Z.~Liu, Y.~Ma, and Y.~Zhang, ``{\it {Millicharged particles at
  electron colliders}},''
  \href{http://dx.doi.org/10.1103/PhysRevD.102.015002}{{\em Phys. Rev. D}
  {\bfseries 102} no.~1, (2020) 015002},
  [\href{http://arxiv.org/abs/1909.06847}{{\ttfamily arXiv:1909.06847}}
  [hep-ph]].

\bibitem{deNiverville:2018hrc}
P.~deNiverville, H.-S. Lee, and M.-S. Seo, ``{\it {Implications of the dark
  axion portal for the muon g\ensuremath{-}2 , B factories, fixed target
  neutrino experiments, and beam dumps}},''
  \href{http://dx.doi.org/10.1103/PhysRevD.98.115011}{{\em Phys. Rev. D}
  {\bfseries 98} no.~11, (2018) 115011},
  [\href{http://arxiv.org/abs/1806.00757}{{\ttfamily arXiv:1806.00757}}
  [hep-ph]].

\bibitem{Zhevlakov:2022vio}
A.~S. Zhevlakov, D.~V. Kirpichnikov, and V.~E. Lyubovitskij, ``{\it
  {Implication of the dark axion portal for the EDM of fermions and dark matter
  probing with NA64e, NA64\ensuremath{\mu}, LDMX, M3, and BaBar}},''
  \href{http://dx.doi.org/10.1103/PhysRevD.106.035018}{{\em Phys. Rev. D}
  {\bfseries 106} no.~3, (2022) 035018},
  [\href{http://arxiv.org/abs/2204.09978}{{\ttfamily arXiv:2204.09978}}
  [hep-ph]].

\bibitem{Jodlowski:2024lab}
K.~Jod\l{}owski, ``{\it {Dark axion portal at $Z$ boson factories}},''
  [\href{http://arxiv.org/abs/2411.19196}{{\ttfamily arXiv:2411.19196}}
  [hep-ph]].

\bibitem{Magill:2018jla}
G.~Magill, R.~Plestid, M.~Pospelov, and Y.-D. Tsai, ``{\it {Dipole Portal to
  Heavy Neutral Leptons}},''
  \href{http://dx.doi.org/10.1103/PhysRevD.98.115015}{{\em Phys. Rev. D}
  {\bfseries 98} no.~11, (2018) 115015},
  [\href{http://arxiv.org/abs/1803.03262}{{\ttfamily arXiv:1803.03262}}
  [hep-ph]].

\bibitem{Zhang:2023nxy}
Y.~Zhang and W.~Liu, ``{\it {Probing active-sterile neutrino transition
  magnetic moments at LEP and CEPC}},''
  \href{http://dx.doi.org/10.1103/PhysRevD.107.095031}{{\em Phys. Rev. D}
  {\bfseries 107} no.~9, (2023) 095031},
  [\href{http://arxiv.org/abs/2301.06050}{{\ttfamily arXiv:2301.06050}}
  [hep-ph]].

\bibitem{Izaguirre:2015zva}
E.~Izaguirre, G.~Krnjaic, and B.~Shuve, ``{\it {Discovering Inelastic
  Thermal-Relic Dark Matter at Colliders}},''
  \href{http://dx.doi.org/10.1103/PhysRevD.93.063523}{{\em Phys. Rev. D}
  {\bfseries 93} no.~6, (2016) 063523},
  [\href{http://arxiv.org/abs/1508.03050}{{\ttfamily arXiv:1508.03050}}
  [hep-ph]].

\bibitem{Dienes:2023uve}
K.~R. Dienes, J.~L. Feng, M.~Fieg, F.~Huang, S.~J. Lee, and B.~Thomas, ``{\it
  {Extending the discovery potential for inelastic-dipole dark matter with
  FASER}},'' \href{http://dx.doi.org/10.1103/PhysRevD.107.115006}{{\em Phys.
  Rev. D} {\bfseries 107} no.~11, (2023) 115006},
  [\href{http://arxiv.org/abs/2301.05252}{{\ttfamily arXiv:2301.05252}}
  [hep-ph]].

\bibitem{Barducci:2024nvd}
D.~Barducci, E.~Bertuzzo, M.~Taoso, C.~A. Ternes, and C.~Toni, ``{\it
  {Illuminating the dark: mono-\ensuremath{\gamma} signals at NA62}},''
  \href{http://dx.doi.org/10.1007/JHEP10(2024)016}{{\em JHEP} {\bfseries 10}
  (2024) 016}, [\href{http://arxiv.org/abs/2406.17599}{{\ttfamily
  arXiv:2406.17599}} [hep-ph]].

\bibitem{OPAL:1990xgw}
{\bfseries OPAL} Collaboration, M.~Z. Akrawy {\em et~al.}, ``{\it {A Direct
  search for neutralino production at LEP}},''
  \href{http://dx.doi.org/10.1016/0370-2693(90)90041-4}{{\em Phys. Lett. B}
  {\bfseries 248} (1990) 211--219}.

\bibitem{CEPCStudyGroup:2025kmw}
{\bfseries CEPC Study Group} Collaboration, S.~P. Adhya {\em et~al.}, ``{\it
  {CEPC Technical Design Report - Reference Detector}},''
  [\href{http://arxiv.org/abs/2510.05260}{{\ttfamily arXiv:2510.05260}}
  [hep-ex]].

\bibitem{CEPCStudyGroup:2023quu}
{\bfseries CEPC Study Group} Collaboration, W.~Abdallah {\em et~al.}, ``{\it
  {CEPC Technical Design Report: Accelerator}},''
  \href{http://dx.doi.org/10.1007/s41605-024-00463-y}{{\em Radiat. Detect.
  Technol. Methods} {\bfseries 8} no.~1, (2024) 1--1105},
  [\href{http://arxiv.org/abs/2312.14363}{{\ttfamily arXiv:2312.14363}}
  [physics.acc-ph]]. [Erratum: Radiat.Detect.Technol.Methods 9, 184--192
  (2025)].

\bibitem{CEPCStudyGroup:2018rmc}
{\bfseries CEPC Study Group} Collaboration, ``{\it {CEPC Conceptual Design
  Report: Volume 1 - Accelerator}},''
  [\href{http://arxiv.org/abs/1809.00285}{{\ttfamily arXiv:1809.00285}}
  [physics.acc-ph]].

\bibitem{CEPCStudyGroup:2018ghi}
{\bfseries CEPC Study Group} Collaboration, M.~Dong {\em et~al.}, ``{\it {CEPC
  Conceptual Design Report: Volume 2 - Physics {\&} Detector}},''
  [\href{http://arxiv.org/abs/1811.10545}{{\ttfamily arXiv:1811.10545}}
  [hep-ex]].

\bibitem{FCC:2018evy}
{\bfseries FCC} Collaboration, A.~Abada {\em et~al.}, ``{\it {FCC-ee: The
  Lepton Collider}: {Future Circular Collider Conceptual Design Report Volume
  2}},'' \href{http://dx.doi.org/10.1140/epjst/e2019-900045-4}{{\em Eur. Phys.
  J. ST} {\bfseries 228} no.~2, (2019) 261--623}.

\bibitem{Agapov:2022bhm}
I.~Agapov {\em et~al.}, ``{\it {Future Circular Lepton Collider FCC-ee:
  Overview and Status}},'' in {\em {Snowmass 2021}}.
\newblock 3, 2022.
\newblock [\href{http://arxiv.org/abs/2203.08310}{{\ttfamily arXiv:2203.08310}}
  [physics.acc-ph]].

\bibitem{ATLAS:2013etx}
{\bfseries ATLAS} Collaboration, G.~Aad {\em et~al.}, ``{\it {Search for
  nonpointing photons in the diphoton and $E^{miss}_T$ final state in
  $\sqrt{s}$=7 TeV proton-proton collisions using the ATLAS detector}},''
  \href{http://dx.doi.org/10.1103/PhysRevD.88.012001}{{\em Phys. Rev. D}
  {\bfseries 88} no.~1, (2013) 012001},
  [\href{http://arxiv.org/abs/1304.6310}{{\ttfamily arXiv:1304.6310}}
  [hep-ex]].

\bibitem{Duarte:2023tdw}
L.~Duarte, J.~Jones-P\'erez, and C.~Manrique-Chavil, ``{\it {Bounding the
  Dimension-5 Seesaw Portal with non-pointing photon searches}},''
  \href{http://dx.doi.org/10.1007/JHEP04(2024)133}{{\em JHEP} {\bfseries 04}
  (2024) 133}, [\href{http://arxiv.org/abs/2311.17989}{{\ttfamily
  arXiv:2311.17989}} [hep-ph]].

\bibitem{Beltran:2024twr}
R.~Beltr\'an, P.~D. Bolton, F.~F. Deppisch, C.~Hati, and M.~Hirsch, ``{\it
  {Probing heavy neutrino magnetic moments at the LHC using long-lived particle
  searches}},'' \href{http://dx.doi.org/10.1007/JHEP07(2024)153}{{\em JHEP}
  {\bfseries 07} (2024) 153},
  [\href{http://arxiv.org/abs/2405.08877}{{\ttfamily arXiv:2405.08877}}
  [hep-ph]].

\bibitem{PandoraPFA2009}
M.~A. Thomson, ``{\it {Particle Flow Calorimetry and the PandoraPFA
  Algorithm}},'' \href{http://dx.doi.org/10.1016/j.nima.2009.09.009}{{\em Nucl.
  Instrum. Meth. A} {\bfseries 611} (2009) 25--40},
  [\href{http://arxiv.org/abs/0907.3577}{{\ttfamily arXiv:0907.3577}}
  [physics.ins-det]].

\bibitem{Ruan:2013rkk}
M.~Ruan and H.~Videau, ``{\it {Arbor, a new approach of the Particle Flow
  Algorithm}},'' in {\em {International Conference on Calorimetry for the High
  Energy Frontier}}, pp.~316--324.
\newblock 2013.
\newblock [\href{http://arxiv.org/abs/1403.4784}{{\ttfamily arXiv:1403.4784}}
  [physics.ins-det]].

\bibitem{Ruan:2018yrh}
M.~Ruan {\em et~al.}, ``{\it {Reconstruction of physics objects at the Circular
  Electron Positron Collider with Arbor}},''
  \href{http://dx.doi.org/10.1140/epjc/s10052-018-5876-z}{{\em Eur. Phys. J. C}
  {\bfseries 78} no.~5, (2018) 426},
  [\href{http://arxiv.org/abs/1806.04879}{{\ttfamily arXiv:1806.04879}}
  [hep-ex]].

\bibitem{Grupen:2008zz}
C.~Grupen and B.~Schwartz, {\em {Particle detectors}}.
\newblock Cambridge Univ. Pr., Cambridge, UK, 2008.

\bibitem{Kaneta:2016wvf}
K.~Kaneta, H.-S. Lee, and S.~Yun, ``{\it {Portal Connecting Dark Photons and
  Axions}},'' \href{http://dx.doi.org/10.1103/PhysRevLett.118.101802}{{\em
  Phys. Rev. Lett.} {\bfseries 118} no.~10, (2017) 101802},
  [\href{http://arxiv.org/abs/1611.01466}{{\ttfamily arXiv:1611.01466}}
  [hep-ph]].

\bibitem{delAguila:2008ir}
F.~del Aguila, S.~Bar-Shalom, A.~Soni, and J.~Wudka, ``{\it {Heavy Majorana
  Neutrinos in the Effective Lagrangian Description: Application to Hadron
  Colliders}},'' \href{http://dx.doi.org/10.1016/j.physletb.2008.11.031}{{\em
  Phys. Lett. B} {\bfseries 670} (2009) 399--402},
  [\href{http://arxiv.org/abs/0806.0876}{{\ttfamily arXiv:0806.0876}}
  [hep-ph]].

\bibitem{Liao:2016qyd}
Y.~Liao and X.-D. Ma, ``{\it {Operators up to Dimension Seven in Standard Model
  Effective Field Theory Extended with Sterile Neutrinos}},''
  \href{http://dx.doi.org/10.1103/PhysRevD.96.015012}{{\em Phys. Rev. D}
  {\bfseries 96} no.~1, (2017) 015012},
  [\href{http://arxiv.org/abs/1612.04527}{{\ttfamily arXiv:1612.04527}}
  [hep-ph]].

\bibitem{Alwall:2014hca}
J.~Alwall, R.~Frederix, S.~Frixione, V.~Hirschi, F.~Maltoni, O.~Mattelaer,
  H.~S. Shao, T.~Stelzer, P.~Torrielli, and M.~Zaro, ``{\it {The automated
  computation of tree-level and next-to-leading order differential cross
  sections, and their matching to parton shower simulations}},''
  \href{http://dx.doi.org/10.1007/JHEP07(2014)079}{{\em JHEP} {\bfseries 07}
  (2014) 079}, [\href{http://arxiv.org/abs/1405.0301}{{\ttfamily
  arXiv:1405.0301}} [hep-ph]].

\bibitem{FASER:2019dxq}
{\bfseries FASER} Collaboration, H.~Abreu {\em et~al.}, ``{\it {Detecting and
  Studying High-Energy Collider Neutrinos with FASER at the LHC}},''
  \href{http://dx.doi.org/10.1140/epjc/s10052-020-7631-5}{{\em Eur. Phys. J. C}
  {\bfseries 80} no.~1, (2020) 61},
  [\href{http://arxiv.org/abs/1908.02310}{{\ttfamily arXiv:1908.02310}}
  [hep-ex]].

\bibitem{Bierlich:2022pfr}
C.~Bierlich {\em et~al.}, ``{\it {A comprehensive guide to the physics and
  usage of PYTHIA 8.3}}''
  \href{http://dx.doi.org/10.21468/SciPostPhysCodeb.8}{{\em SciPost Phys.
  Codeb.} {\bfseries 2022} (2022) 8},
  [\href{http://arxiv.org/abs/2203.11601}{{\ttfamily arXiv:2203.11601}}
  [hep-ph]].

\end{thebibliography}\endgroup
\bibliographystyle{utphys}

\newpage
\onecolumngrid

\newpage
\widetext
\clearpage
\begin{center}
 \textbf{Supplemental Materials}
\end{center} 

\subsection{A. Sensitivities to the neutrino dipole portal}
\begin{figure}[htbp]
\centering
\includegraphics[width=0.4 \textwidth]{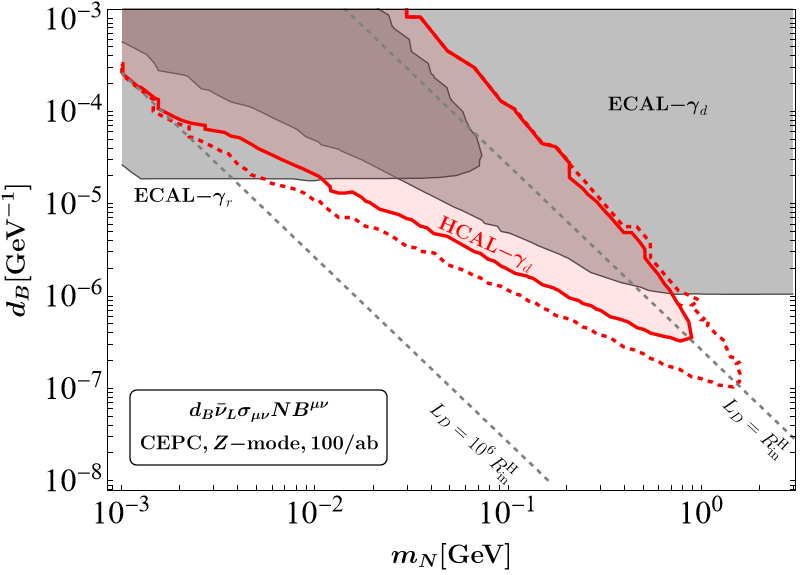}
\caption{
The expected $2\sigma$ sensitivities to the neutrino dipole portal for the CEPC $Z$-mode for various mono-photon signatures, including the ECAL-$\gamma_d$, the ECAL-$\gamma_r$, and the HCAL-$\gamma_d$ signatures.
For the HCAL-$\gamma_d$ signature, the solid (dashed) line represents the conservative (optimistic) estimate with $N_b = 10^4$ ($10^2$).
The gray dashed lines indicate the parameter space with $L_D = R_{\rm in}^H$ and $L_D = 10^6\, R_{\rm in}^H$ for the sterile neutrino in the laboratory frame.
}
\label{fig:con-ND}
\end{figure}

Here we present the sensitivities to the neutrino dipole portal from three mono-photon signatures.
The dimension-six $\nu$SMEFT operator $(\bar{L} \sigma_{\mu\nu} N) \tilde{H} B^{\mu\nu}$ \cite{delAguila:2008ir,Liao:2016qyd} can induce the neutrino dipole portal operator  \cite{Magill:2018jla,Zhang:2023nxy},
\begin{equation}
\mathcal{O}_{\nu N B} \equiv d_B \bar{\nu}_L \sigma_{\mu\nu} N B^{\mu\nu},
\end{equation}
where $\nu_L$ denotes the SM left-handed neutrino, $N$ the sterile right-handed neutrino, $L$ the SM lepton doublet, $H$ the Higgs doublet, $B_{\mu\nu}$ the field-strength tensor of the SM hypercharge field $B$, and $d_B$ the neutrino dipole moment with the mass dimension of $-1$.
Here we assume that only one neutrino flavor participates in the interaction for simplicity.  
Then, the decay width of $N \to \nu \gamma$ is given by \cite{Magill:2018jla,Zhang:2023nxy}
\begin{equation}
    \Gamma_{N\to \nu \gamma} = \frac{1}{4\pi} d_B^2 c_W^2 m_N^3,
\end{equation}
where $m_N$ is the mass of the sterile neutrino and  $c_W = \cos\theta_W$, with $\theta_W$ denoting the Weinberg angle.
For the CEPC $Z$-mode, Fig.~\ref{fig:con-ND} shows the $2\sigma$ sensitivities to the neutrino dipole portal from three mono-photon signatures, as a function of $m_N$.
For the HCAL-$\gamma_d$ signature,
we adopt the five-layer configuration and estimate the conservative (optimistic) sensitivities with $N_b=10^4$ ($N_b=10^2$) for GS-HCAL.
Similar to the dark axion portal case, the HCAL-$\gamma_d$ signature exhibits excellent sensitivity when the decay length of the sterile neutrino lies between $R_{\rm in}^H$ and $10^6 R_{\rm in}^H$.

\subsection{B. The SM backgrounds for the HCAL-$\gamma_d$ signature}
The SM backgrounds for the HCAL-$\gamma_d$ signature arise from neutral particles produced at the primary vertex in the absence of other visible energetic particles.
These neutral particles then evade detection in the inner tracker and ECAL, depositing energy only in the HCAL barrel.
We estimate in this section the dominant contributions from photon-, hadron-, and neutrino-induced processes.

\textbf{Photon-induced backgrounds:} 
Photons generated by $e^+ e^- \to \nu \bar{\nu} \gamma$, accompanied by invisible neutrinos, can contribute to the backgrounds if the photon penetrates the ECAL barrel region.
Our \textsc{MadGraph}~\cite{Alwall:2014hca} simulation yields $N^\gamma_{\rm EB} = 1.34 \times 10^9$ photons within the HCAL barrel acceptance and above the photon energy threshold of the five-layer requirement for a luminosity of 100 ab$^{-1}$. 
The probability for these photons to penetrate the ECAL barrel and initiate showers in the HCAL is given by the exponential attenuation law:
$
P^{\slashed{\gamma}}_{\text{EB}} = e^{-7/9 \times 24} = 7.8\times10^{-9},
$
where the factor $7/9$ accounts for the ratio of the radiation lengths of electrons to photons, and $24$ corresponds to the ECAL thickness in the unit of $X_0$. This yields approximately 10 photon-induced background events.
Crucially, the HCAL barrel's 48-layer structure provides additional rejection capability. Using the first three layers (4.3$X_0$) as a veto reduces these backgrounds to negligible levels with only minimal signal efficiency loss, as genuine displaced photons from LLP decays typically shower deeper in the HCAL volume.

\textbf{Neutrino-induced backgrounds:} Electron neutrino pairs from $e^+e^- \to \nu_e\bar{\nu}_e$ processes represent another potential background source. Our simulation identifies $N_{\nu\bar{\nu}} = 4.1 \times 10^9$ electron neutrino pairs within the HCAL barrel acceptance, each carrying $E_\nu = 45.5$ GeV.
These neutrinos can easily penetrate the ECAL and contribute to background events if they undergo 
the charge current interaction in the HCAL barrel, producing an electromagnetic shower from the outgoing electron or positron.
The number of background events from neutrino interactions can be conservatively estimated using the empirical relation~\cite{FASER:2019dxq}:
\begin{equation}
N_{\rm BG}^\nu = 2 \times N_{\nu\bar{\nu}} \times 4 \times 10^{-13} \times \frac{E_\nu}{\rm GeV} \times \frac{L_{\rm HB}}{\rm m} \times \frac{\rho_{\rm HB}}{\rho_{\rm H_2O}},
\end{equation}
where $L_{\rm HB} = 1.32$ m is the HCAL barrel radial length and $\rho_{\rm HB} \approx 5.0$ g/cm$^3$ is its average density. This yields approximately 1.2 events, confirming that neutrino-induced backgrounds are negligible.

\textbf{Hadron-induced backgrounds:} The most significant backgrounds arise from neutral hadrons (primarily neutrons and $K_L$ mesons) produced in hadronic events where all other particles escape detection along the beam direction. To quantify this background, we generated $10^8$ parton-level three-jet events using \textsc{MadGraph} with selection criteria $p_T^{j_1} > 1$ GeV, $p_T^{j_2,j_3} < 10$ GeV, and $\Delta R_{2j} > 0.4$, corresponding to a cross section of $\sigma_{3j} = 3300$ pb.
These events were processed through \textsc{Pythia}~\cite{Bierlich:2022pfr} for showering and hadronization. After applying vetoes on any charged particle with $p_T > 0.1$ GeV or photon with $E_\gamma > 0.1$ GeV within detector acceptance, we identified events with only one neutral hadron (neutron or $K_L$) in the HCAL barrel.
The probability for neutrons/$K_L$ to penetrate the ECAL is $P^{\slashed{n}}_{\text{EB}} = e^{-1.35} = 0.26$, where 1.35 accounts for the length of the ECAL barrel in the unit of $\lambda_I$.

With only one such event found in our simulation sample, the hadron-induced background is estimated as:
\begin{equation}
N^{n, K_L}_{\rm BG} \simeq \frac{1}{10^8} \times \sigma_{3j} \times \mathcal{L} \times P^{\slashed{n}}_{\text{EB}} = 858 \text{ events}.
\end{equation}

The distinct shower profiles of photons versus neutral hadrons enable powerful background discrimination through the Particle Flow Algorithm. Photon-induced electromagnetic showers exhibit characteristic longitudinal and lateral development patterns that differ significantly from hadronic interactions.
An optimized photon identification efficiency of approximately 10\%  can reduce the hadron-induced backgrounds to fewer than 100 events.

\subsection{C. The SM backgrounds for the ECAL-$\gamma_d$ and ECAL-$\gamma_r$ signatures}
Unlike charged particles such as electrons and muons, which leave distinct tracks in the tracking system, neutral photons only produce clusters in the calorimeter.
Consequently, in the absence of a high-resolution directional ECAL, both the ECAL-$\gamma_d$ and ECAL-$\gamma_r$ signatures manifest as prompt mono-photon events in the ECAL and are subject to similar SM backgrounds.

The SM backgrounds for ECAL mono-photon events at CEPC have been extensively discussed in Ref.~\cite{Liu:2019ogn}, including both the irreducible and reducible backgrounds. 
For the CEPC $Z$-mode, with the center-of-mass energy around the $Z$ pole, 
a substantial irreducible background arises from $e^+ e^- \to \nu \bar{\nu} \gamma$.
Applying preselection cuts of $E_\gamma > 1~\rm{GeV}$ and $|\cos\theta_\gamma| < |\cos\theta_c|$, there are $1.25 \times 10^{10}$ $e^+ e^- \to \nu \bar{\nu} \gamma$ events with the luminosity of 100/ab, where $\theta_\gamma$ ($\theta_c$) is the polar angle of the photon (ECAL edges) along the beam direction.
The reducible backgrounds arise from $e^+ e^- \to \slashed{\gamma} \slashed{\gamma} \gamma$ and $e^+ e^- \to \slashed{\ell}^+ \slashed{\ell}^- \gamma$ with $\ell = e, \mu$, where the slashed particles are outside the detector acceptance.
With a highly hermetic detector, all such reducible backgrounds can be completely removed by a $\theta_\gamma$-dependent cut on the photon energy~\cite{Liu:2019ogn, Liang:2019zkb}: 
\begin{equation} 
E_\gamma > E_B^m(\theta_\gamma) \equiv \sqrt{s} \left[ 1+ \frac{\sin\theta_\gamma}{\sin \theta_c} \right]^{-1}. 
\label{eq:bGB-cut}
\end{equation}
With the cut of $E_\gamma > E_B^m(\theta_\gamma)$ and the preselection cuts, the number of irreducible background events is $4.22 \times 10^7$ for a luminosity of 100/ab.

In summary, for the two ECAL-based mono-photon signatures, we apply the cuts of $E_\gamma > 1~\rm GeV$, $|\cos\theta_\gamma| < |\cos\theta_c|$, and $E_\gamma > E_B^m(\theta_\gamma)$. 
These cuts yield a total SM background of $4.22 \times 10^7$ for a luminosity of 100/ab.

\end{document}